\documentstyle[aps,prb,epsf]{revtex}
\input amssym.def
\newtheorem{thm}{Theorem}[section]

\newtheorem{lem}[thm]{Lemma}

\newtheorem{remark}[thm]{Remark}

\newcommand{\bbR}{\Bbb R }

\newcommand{\calB}{\cal B }
\newcommand{\calC}{\cal C}

\newcommand{\calS}{\cal S}

\newcommand{\calZ}{\cal Z}

\newcommand{\grad}{\mbox{\rm grad}}
\newcommand{\cm}{\mbox{\rm cm}}
\newcommand{\rad}{\mbox{\rm rad}}
\newcommand{\Hz}{\mbox{\rm Hz}}
\newcommand{\lb}{\label}

\begin{document}
\draft
\author{
C. Chicone\thanks{ Department of  Mathematics,
University of Missouri, Columbia, MO 65211.
Supported by the NSF grant  DMS-9303767.} ,\quad
B. Mashhoon\thanks{ Department of  Physics and Astronomy,
University of Missouri, Columbia, MO 65211.
}\quad
and D. G. Retzloff\thanks{ Department of Chemical Engineering,
University of Missouri, Columbia, MO 65211.}
}
\date{\today}

\title{On the Ionization of a Keplerian Binary System by Periodic Gravitational
Radiation}

\maketitle

\begin{abstract}  The gravitational ionization of a Keplerian binary system via
normally incident periodic gravitational radiation of definite helicity is
discussed.  The periodic
orbits of the planar tidal equation are investigated on the basis of degenerate
continuation theory.  The relevance of the Kolmogorov-Arnold-Moser theory to the
question of gravitational ionization is elucidated, and it is conjectured that
the process of ionization is closely related to the Arnold diffusion of the
perturbed system.
\end{abstract}


\section{Introduction}
\lb{intros}
In a recent paper \cite{cmr}, we considered the long-term nonlinear
perturbations of Keplerian orbits by incident gravitational waves of wavelengths
much larger than the size of the system.  In particular, we studied the periodic
orbits of the perturbed system using the 
methods developed in Ref.\ \onlinecite{ccc}.  The
existence of periodic orbits indicates the possibility of balance in the
exchange of energy between the binary and the external radiation field.  Thus
gravitational ionization does not occur for such orbits.  The issue of
gravitational ionization is interesting as it involves the transport of energy
by gravitational radiation and is analogous to the corresponding phenomenon that
is well known in the electromagnetic context.

The theoretical investigation of the interaction
of a binary system with the gravitational radiation field
reveals subtle phenomena that are further studied in this paper.
In particular, the absorption of gravitational radiation energy
by the binary is not unidirectional in general. That is, the orbital energy of
a binary immersed in a gravitational radiation field does not in general
increase monotonically with time. 
On the other hand,
the emission of gravitational radiation by the binary is expected to be
accompanied by the monotonic decrease of the orbital energy of the system.
In absorption, however, the incident wave can deposit energy into the
orbit during one time interval and remove energy from the orbit during
another time interval. A periodic orbit would result---even when the emission
of the radiation by the binary is ignored---if after a certain time the
net flow of energy between the incident wave and the binary is zero.

Let us imagine, for the sake of simplicity, that gravitational radiation is
incident on a Newtonian binary system consisting of a massive body of mass
$M_0$  at the origin of inertial coordinates and a particle of test mass $m_0 <<
M_0$ that revolves around it in the $(x,y)$-plane.  The dynamical equation in
this case is of the general form \cite{mashoon1,mashoon2}
\begin{equation}\label{BasicEq1}
\frac{d^2x^i}{dt^2} + \frac{kx^i}{r^3} + \epsilon {\cal K}_{ij}(t) x^j = 0 ,
\end{equation}
where $k = G_0(M_0+m_0)$, $\epsilon$, 
$0 < \epsilon << 1$, is the perturbation parameter and
$\epsilon {\cal K}_{ij}$ is the tidal matrix associated with the incident
gravitational waves.  Here $\cal K$ is symmetric and traceless, and is related
to the gravitational perturbation of the Minkowski space-time by
\begin{equation}\label{kijeq2}
{\cal K}_{ij}(t) = - \frac{1}{2} \frac{\partial^2 \chi_{ij}}{\partial t^2}
(t,{\bf 0}),
\end{equation}
where $g_{\mu\nu} = \eta_{\mu\nu} + \epsilon \chi_{\mu\nu}$.  Here we employ
the
transverse-traceless gauge for gravitational radiation,
i.e. $\chi_{0\mu} = 0$,
$\chi_{ij}$ is traceless and $\partial_j \chi_{ij} = 0$;
moreover,  $\chi_{ij}$
is a solution of the wave equation
$\Box \chi_{ij} = 0$.
Thus the
gravitational radiation field can be expressed as a Fourier sum of
monochromatic waves of frequency $\Omega_{\star}$ and wave vector
${\bf K}_{\star}$,  
$c|{\bf K}_{\star}| = \Omega_{\star}$,
\begin{equation}\label{chi1}
\chi_{ij}(t, {\bf x}) = \mbox{\rm Re} \sum_{{\bf K}_{\star}} \widehat{\chi}_{ij}
({\bf
K}_{\star}) \exp{(i{\bf K}_{\star}\cdot {\bf x} - i\Omega_{\star} t)},
\end{equation}
where $\widehat{\chi}_{ij}$ is symmetric, traceless and $\widehat{\chi}_{ij}K^j_{\star}
= 0$.  The summation in (\ref{chi1}) extends over all waves with $2\pi
c/\Omega_{\star}$ much larger than the average orbital radius.  Equation
(\ref{BasicEq1}) contains only the essential physics of the interaction of
long-wavelength gravitational radiation with a Newtonian binary system; in fact,
relativistic (i.e.\ post-Keplerian) effects in the binary are totally
neglected.  In particular, the emission of gravitational waves is ignored.  The
motivation for our treatment as well as its limitations is presented in detail
in our recent work \cite{cmr}.

The incident wave exchanges energy and angular momentum with the binary orbit
but not linear momentum in the {\em quadrupole} approximation under
consideration here \cite{cmr}.  This is in exact analogy with the
electromagnetic problem of the interaction of an electromagnetic wave with an
atom in the {\it dipole} approximation.

A simple linear perturbation treatment of (\ref{BasicEq1}) has revealed the
possibility of the existence of resonances at $\Omega_{\star} = m \omega$,
$m=1,2,3,\cdots$, where $\omega$ is the Keplerian frequency of the unperturbed
elliptical orbit.  Moreover, in this analysis secular terms appear that lead to
the breakdown of the linear theory over time \cite{mashoon2}.
Thus linear perturbation theory is inappropriate for the
investigation of periodic orbits of the perturbed system, since a periodic
orbit is expected to persist forever beyond a certain point in time.

In the first treatment of the nonlinear case \cite{cmr}, we considered a single
monochromatic plane wave of frequency $\Omega_{\star}$ that was normally
incident on a Keplerian orbit of frequency $\omega$.  We found that in the
generic case, certain orbits satisfying the resonance condition $\Omega_{\star}
= m \omega$, $m=1,2,3,\cdots$, could be continued to periodic orbits of the
nonlinear system.  The existence of periodic solutions of (\ref{BasicEq1})
demonstrates that ionization does not always occur; in fact, in a periodic orbit
the energy exchange with the radiation field must be steady without any net
flow.  In addition, we found that for incident circularly polarized radiation of
definite helicity the rotation of the inertial coordinates by frequency
$\Omega_{\star}/2$ rendered the dynamical equations autonomous.  The invariance
of this autonomous system under time translation implies the existence of an
energy
integral in the rotating frame.  In this case, the Kolmogorov-Arnold-Moser (KAM)
theory implies that for sufficiently small $\epsilon$ ionization can never occur
in this system regardless of the magnitude of $\Omega_{\star}/\omega$.  To
understand intuitively how this could come about, it should be pointed out
that a binary system can gain or lose energy as it interacts with an incident
gravitational wave.  The situation in absorption is in contrast to the
emission of gravitational waves by a binary.  In the latter situation, the
binary is expected to lose energy monotonically; in fact, this is consistent
with the
observed rate of inward spiraling of the Hulse-Taylor binary pulsar
\cite{hulse,taylor}.  It follows that the reciprocity between emission and
absorption of gravitational waves does not hold in general.  This notion of
reciprocity is valid in some other situations, however.  For instance, a
Keplerian ellipse of frequency $\omega$ emits gravitational radiation of
frequency $m\omega, m = 1,2,3,\cdots$, which corresponds to the resonance
condition for absorption.

Let us now consider a {\em general} periodic gravitational wave of period
$2\pi/\Omega_{\star}$ that is normally incident on the binary system.  The
existence of certain periodic solutions of the perturbed system may be expected
on general grounds.  It would therefore be more interesting to investigate the
interaction of circularly polarized gravitational radiation with the binary
system and to determine the stability of the resulting autonomous system under
periodic perturbations.  That is, a periodic wave may be expressed as a Fourier
sum of components with frequencies $n\Omega_{\star}$, $n = 1,2,3,\cdots$.  For a
single component of definite helicity, the transformation to the corresponding
rotating frame would essentially remove the dependence of this perturbation upon
time and for $\epsilon$ below a certain limit the orbit would remain forever
bounded even though there is a steady flow of incident gravitational radiation
energy  in the inertial frame.  However, the time-dependence of the other
Fourier components would not disappear in the rotating frame, and we would like
to study the influence of these components on the ionization of the system.
Though we develop methods that are applicable to a general periodic
perturbation, we restrict our attention to a tractable problem for the sake of
simplicity.

In this paper, we consider a superposition of several harmonics in the
perturbing function; clearly, the response of the system is not a superposition
of the individual responses as a consequence of the intrinsic nonlinearity of
the problem under consideration here.  Specifically, we showed in 
Ref.\ \onlinecite{cmr}
that for a normally incident circularly polarized monochromatic plane wave the
motion is restricted to the $(x,y)$-plane and that a transformation to the
uniformly rotating coordinate system in this plane with half the wave's
frequency would result in an autonomous system for the equation of motion to
which the  KAM theory can be applied.  It follows from the KAM theory that for
sufficiently small $\epsilon$ the motion is confined and ionization does not
occur.  We wish to explore the sensitivity of this interesting result to the
particular form of the incident wave.  Therefore, we consider here a principal
right circularly polarized wave of frequency  $\Omega_{\star} = 2 \Omega$ that
is slightly modified by the presence of similar components of frequencies
$\Omega$ and $3\Omega$ as follows:
\begin{eqnarray}\label{chimatrix}
\chi_{11}(t, {\bf 0}) & = & \cos{2\Omega t} + 2\delta \left[ (\alpha - \beta)
\cos{\Omega t}
+ \frac{1}{9}(\alpha + \beta) \cos{3 \Omega t} \right], \nonumber \\
\chi_{12}(t, {\bf 0}) & = & \sin{2\Omega t} + 2 \delta \left[(\alpha - \beta)
\sin{\Omega t}
+ \frac{1}{9}(\alpha + \beta) \sin{3\Omega t} \right],
\end{eqnarray}
$\chi_{13} = \chi_{23} = \chi_{33} =  0$. The other components of $\chi$ follow
from the fact that $\chi$ is a  symmetric traceless matrix.  
Here $\delta$, $0 < \delta << 1$, is a new perturbation parameter 
that determines the relative
strength of the extra secondary components compared to the primary Fourier
component of the normally incident radiation.  Moreover, $\alpha$ and $\beta$
are constant amplitudes of the order of unity, and the other numerical
coefficients have been introduced for the sake of simplicity.

The plan of this paper is as follows: In Section~\ref{desec}, 
we present the  basic
equations for a Hamiltonian description of the perturbed orbit in terms of
Delaunay variables.  Sections~\ref{bifsec}--\ref{secpo}
are devoted to a development of degenerate
continuation theory that is necessary for the identification of periodic orbits
of the nonlinear problem via the methods and ideas that are originally due to
Poincar\'{e}.  The existence of periodic orbits, described in 
Section~\ref{secpo},
demonstrates that 
a state of equilibrium can be established
between the wave and the binary such that ionization does not occur; in
fact, the net flow of energy vanishes in this case.  To apply the KAM theory to
our problem, it is best to transform (\ref{BasicEq1}) to a uniformly rotating
frame as in Section~\ref{rfsec}.  
It follows from the description of the nonlinear system
in this reference frame that Arnold diffusion is expected for $\delta > 0$, even
for sufficiently small $\epsilon$.  Numerical experiments described in 
Section~\ref{nesec}
tend to corroborate the conjecture that gravitational ionization is tantamount
to Arnold diffusion in this system.  For background material, this paper relies
heavily on our previous detailed treatment of the nonlinear problem for the case
where the incident wave is essentially a simple monochromatic Fourier component
\cite{cmr}; however, we have attempted to present sufficient detail here in
order to render the present paper essentially self-contained.

\section{Hamiltonian Description in Delaunay Elements}
\lb{desec}
Using (\ref{BasicEq1}), (\ref{kijeq2}) and (\ref{chimatrix}), we can
write the associated Hamiltonian for this system as the sum of the Kepler
Hamiltonian and the quadrupole perturbation given by $\frac{1}{2} \epsilon{\cal
K}_{ij}x^ix^j$.  The motion is taken to be in the $(x, y)$-plane, since the
radiation is transverse and normally incident on the orbital plane; therefore,
polar coordinates are convenient.  Defining $\phi(t)$ and $\psi(t)$ as
\begin{eqnarray}
\phi(t) & = &  \Omega^2\{\cos{2\Omega t} + \frac{\delta}{2}[ (\alpha-\beta)
\cos{\Omega t}+(\alpha+\beta) \cos{3\Omega t}]\}, \nonumber \\
\psi(t) & = & \Omega^2\{\sin{2\Omega t} + \frac{\delta}{2}[ (\alpha-\beta)
\sin{\Omega t}+(\alpha+\beta) \sin{3\Omega t}]\},
\end{eqnarray}
the Hamiltonian in polar coordinates may be written as
\begin{equation}\label{Ham1}
{\cal H}  =  \frac{1}{2} \left(p_r^2 + \frac{p_{\theta}^2}{r^2}\right) -
\frac{k}{r} + \epsilon r^2[\phi(t)\cos{2\theta}+\psi(t)\sin{2\theta}].
\end{equation}

To express the Hamiltonian in a form particularly suitable for the analysis of
orbital dynamics, we transform to the Delaunay elements $(L,G,\ell,g)$.
To this end, consider the bounded motion of the test particle according to the
Hamiltonian (\ref{Ham1}).  At each instant of time, the particle can be
described as belonging to an osculating Keplerian ellipse; that is, the
perturbed motion passes through an infinite sequence of osculating ellipses in
the course of time.  Each such ellipse is described by the unperturbed
Hamiltonian
\[ H = \frac{1}{2}\left( p_r^2 + \frac{p_{\theta}^2}{r^2}\right) - \frac{k}{r},
\] where we consider only bounded motions with $E = H(p_r, p_{\theta}, r,
\theta) < 0$.  Let us now consider the canonical transformation from $(p_r,
p_{\theta}, r, \theta)$ to variables intrinsic to the ellipse.  Thus we define
action variables
\[L := \left( \frac{-k^2}{2E}\right)^{\frac{1}{2}},\: \: \:  \mbox{\rm and}\: \:
\: \:  G := p_{\theta},\] which correspond to an osculating Keplerian ellipse
with semimajor axis $a$, $a = L^2/k$, and eccentricity $e$,
\[e = \left(1-\frac{G^2}{L^2}\right)^{\frac{1}{2}},\] 
such that $0 \leq e < 1$.
The equation of this ellipse is given by
\[r = a (1-e \cos{\widehat{u}}), \: \: \: \:\mbox{\rm or} \: \: \: \:  r = a
\frac{1-e^2}{1+e\cos{v}} \: \: ,\] where $\widehat{u}$ is the eccentric anomaly and
$v$ is the true anomaly.  The new canonical angle variables $\ell$ and $g$ are
then given by
\[\ell = \widehat{u}-e\sin{\widehat{u}},\: \: \: \:   g = \theta - v.\]
In the following, we exclude $e = 0$ and focus attention instead on noncircular
elliptical orbits.  The resulting Hamiltonian of the perturbed system in
Delaunay variables is
\begin{equation}\label{Ham2}
{\cal H} = -\frac{k^2}{2L^2} + \epsilon [{\calC}(L,G,\ell,g)\phi(t)+{\cal S}
(L,G,\ell,g)\psi(t)],
\end{equation}
where, in polar coordinates, ${\cal C} = r^2 \cos{2\theta}$ and  ${\cal S}= r^2
\sin{2\theta}$.  In Delaunay elements,  ${\cal C}(L,G,\ell,g)$ and ${\cal
S}(L,G,\ell,g)$ are given by
\begin{eqnarray}\label{CSfs}
{\cal C}(L,G,\ell,g) & = &\frac{5}{2}a^2e^2\cos{2g} \nonumber \nonumber \\
  & & \hspace*{.15in} +a^2\sum_{\nu =1}^\infty (A_\nu(e)\cos{2g}\cos{\nu\ell}
                            -B_\nu(e)\sin{2g}\sin{\nu \ell}),  \nonumber \\
{\cal S}(L,G,\ell,g) & = &\frac{5}{2}a^2e^2\sin{2g} \nonumber \\
  & & \hspace*{.15in} +a^2\sum_{\nu =1}^\infty (A_\nu(e)\sin{2g}\cos{\nu\ell}
                            +B_\nu(e)\cos{2g}\sin{\nu \ell}),
\end{eqnarray}
where
\begin{eqnarray}
A_\nu(e) & = &\frac{4}{\nu^2e^2}
            (2\nu e(1-e^2)J_\nu'(\nu e)-(2-e^2)J_\nu(\nu e)), \nonumber \\
B_\nu(e) & = &-\frac{8}{\nu^2e^2}\sqrt{1-e^2}\,
     (e J_\nu'(\nu e)-\nu (1-e^2)J_\nu(\nu e)). \label{ABessel}
\end{eqnarray}
Here $J_{\nu}$ is the Bessel function of order $\nu$, and a  prime
indicates the derivative of the function with respect to its argument.
The dynamical equations are derived in the usual way from the Hamiltonian
(\ref{Ham2}).   Moreover,  these equations
are given in Delaunay elements by
\begin{eqnarray}\label{EqofMotion}
\dot{L}  & = &  -\epsilon
  \left( \frac{\partial{\cal C}(L,G,\ell,g)}{\partial \ell}\phi(t)
   +\frac{\partial{\cal S}(L,G,\ell,g)}{\partial \ell}\psi(t)\right),\nonumber
\\
\dot{G} & = &  -\epsilon
 \left(
\frac{\partial{\cal C}(L,G,\ell,g)}{\partial g}\phi(t)
  +\frac{\partial{\cal S}(L,G,\ell,g)}{\partial g}\psi(t)\right), \nonumber \\
\dot{\ell} & = & \omega+\epsilon
  \left( \frac{\partial{\cal C}(L,G,\ell,g)}{\partial L}\phi(t)
   +\frac{\partial{\cal S}(L,G,\ell,g)}{\partial L}\psi(t)\right), \nonumber \\
\dot{g} & = &  \epsilon
  \left( \frac{\partial{\cal C}(L,G,\ell,g)}{\partial G}\phi(t)
    +\frac{\partial{\cal S}(L,G,\ell,g)}{\partial G}\psi(t)\right),
\end{eqnarray}
where $\omega$ is the Keplerian frequency of the binary given by  $\omega^2 =
k/a^3$.

\section{Bifurcation Function}
\lb{bifsec}
To establish the continuation (persistence) of periodic orbits of the
Kepler problem to the system (\ref{EqofMotion}), 
we employ a method proposed
in Ref.\ \onlinecite{ccc} and used in Ref.\ \onlinecite{cmr};  
we only state the main ideas here and the
reader is referred to these references for details.

System (\ref{EqofMotion}) has the abstract form
\begin{equation}
\label{bifeq}
\dot{u} = F(u)+\epsilon h(u,t),
\end{equation}
where $u$ is a
coordinate on a manifold $M$ that consists of a cross product of Euclidean
spaces and tori, 
and the function $h$ is periodic with period
$2\pi/\Omega$
in its second argument.
We consider solutions
$t\mapsto u(t,\xi,\epsilon)$ of (\ref{bifeq}) with initial condition
$u(0,\xi,\epsilon) = \xi$, $\xi \in M$, and define the $m$th order Poincar\'{e}
map by ${\cal P}^m(\xi,\epsilon)=u(2\pi m/\Omega,\xi,\epsilon)$.  Fixed points
of this $m$th order Poincar\'{e} map correspond to periodic solutions of
(\ref{bifeq}).  Consider the unperturbed periodic solutions of (\ref{bifeq})
with $\epsilon = 0$.  These correspond to fixed points of the unperturbed $m$th
order Poincar\'{e} map defined by $p^m(\xi)={\cal P}^m(\xi,0)$. Let us
now suppose that
there is a submanifold ${\cal Z}\subset M$ that consists
of fixed points of $p^m$,
and $\zeta \in {\cal Z}$.  
If there is a continuous curve $\epsilon \mapsto
\kappa(\epsilon)$ in $M$ such that $\kappa(0)=\zeta$ and ${\cal
P}^m(\kappa(\epsilon),\epsilon) \equiv \kappa(\epsilon)$, i.e. for each fixed
value of $\epsilon$, $\kappa(\epsilon)$ is a fixed point of the $m$th order
Poincar\'{e} map, then the unperturbed periodic orbit is continuable.  The
continuation is established by the method of Lyapunov-Schmidt reduction to the
Implicit Function Theorem and requires that every vector in $T_{\zeta}M$ that is
tangent to the submanifold ${\cal Z}$ be in the kernel of the infinitesimal
displacement ${\cal D}(\zeta)= Dp^m(\zeta)-I$.  This is equivalent to the
requirement that for each $\zeta\in {\cal Z}$, the dimension of the kernel of
the infinitesimal displacement at $\zeta$ be equal to the dimension of the
manifold $\cal Z$.  The manifold ${\cal Z}$ is called normally nondegenerate if
it satisfies this condition.

Let ${\cal Z}$ be a normally nondegenerate submanifold of $M$ with dimension
$q$; then, the range of the infinitesimal displacement at each point of ${\cal
Z}$ has codimension $q$.  For each $\zeta \in {\cal Z}$, there is a vector
complement $\widehat{{\cal S}}(\zeta)$ to the range of the infinitesimal
displacement of dimension $q$.  We denote the projection of $T_{\zeta}M$ to
$\widehat{{\calS}}(\zeta)$ by $\widehat{s}(\zeta)$.

Let $\zeta\in {\cal Z}$ and consider the curve in the manifold
$M$ defined by $\epsilon\mapsto {\cal P}^m(\zeta,\epsilon)$; 
it passes through $\zeta$ at
$\epsilon=0$ and its tangent vector at $\epsilon=0$ is in $T_{\zeta}M$. 
This tangent vector can be
identified with the partial derivative ${\cal P}^m_\epsilon(\zeta,0)$.
The bifurcation function $\calB$ is defined to be the map
from $\cal Z$ to the complement $\widehat{{\cal S}}$ of the
range of the infinitesimal displacement
\[
{\cal B}(\zeta)=\widehat{s}(\zeta){\cal P}^m_\epsilon(\zeta,0),
\]
so that in local coordinates ${\cal B}:{\bbR}^q\to{\bbR}^q$.
We define $\zeta\in{\cal Z}$ to be a simple zero of the bifurcation function
if ${\cal B}(\zeta)=0$ and the derivative $D{\cal B}(\zeta)$ is
invertible. 
The following continuation theorem is proved in Ref.\ \onlinecite{ccc}.
\begin{thm}\label{conthm}
Let $\cal Z$ denote a normally nondegenerate fixed point
submanifold of $M$ for the system~(\ref{bifeq}). If
$\zeta\in {\cal Z}$ is a simple zero of the corresponding
bifurcation function,
then the unperturbed periodic orbit
of~(\ref{bifeq}) with initial point $\zeta$ is continuable.
\end{thm}

To apply Theorem~\ref{conthm} to the perturbed Kepler problem,
we must compute the partial derivative ${\cal P}_\epsilon^m(\zeta,0)$ 
of the corresponding Poincar\'e map. For the system~(\ref{EqofMotion}), 
the manifold $M$ is the four dimensional Delaunay coordinate
space and the Poincar\'e map is defined as the strobe 
with period $ 2\pi m/\Omega$ where $\Omega$ is
the frequency of the perturbation.
The partial derivative is obtained from the solution 
$t \mapsto W(t)$ of the second variational initial value problem
\[
\dot W=DF(u(t,\zeta,0))W+h(u(t,\zeta,0),t),\quad W(0)=0.
\]
In fact, we have $W(t)=u_\epsilon(t,\zeta,0)$ and 
\[{\cal P}^m_\epsilon(\zeta,0)=W(2\pi m/\Omega).\]

We use Theorem~\ref{conthm} to establish the existence
of periodic orbits for the system~(\ref{EqofMotion}). Of course,
this will establish the existence of periodic orbits for our
model system~(\ref{BasicEq1}), which describes the perturbation of a
Keplerian binary system by a multi-frequency periodic gravitational wave.  
We begin by identifying the normally 
nondegenerate fixed point submanifold of $M$
mentioned in Theorem \ref{conthm}.  
Recall that the frequency of the unperturbed periodic Keplerian 
orbit is $\omega=k^2/L^3$. 
The three-dimensional manifold
\[
{\cal Z}^L:=\{(L,G,\ell,g): m\omega=n\Omega\},
\]
where $m$ and $n$ are relatively prime positive integers,
is a normally nondegenerate submanifold of $M$ (cf.\  Ref.\ \onlinecite{cmr}). 
Furthermore,
the range of the infinitesimal displacement is complemented by the span of
the vectors
\[
\left(
\begin{array}{c}
1  \\
0 \\
0 \\
0
\end{array}
\right),\quad
\left(
\begin{array}{c}
0  \\
1 \\
0 \\
0
\end{array}
\right),\quad
\left(
\begin{array}{c}
0  \\
0 \\
0 \\
1
\end{array}
\right).
\]

The bifurcation function associated with (\ref{EqofMotion})
is the projection of the partial derivative
${\cal P}^m_\epsilon(L,G,\ell,g,0)$  on the manifold
${\cal Z}^L$ onto the complement
of the range of the infinitesimal displacement.
To determine the bifurcation function, we solve the variational initial
value problem
\begin{eqnarray}
\dot L_\epsilon & = & -\frac{\partial {\cal C}}{\partial\ell}
       (L,G,\ell+\omega t,g)\phi(t)
       -\frac{\partial {\cal S}}{\partial \ell}(L,G,\ell + \omega t,g)\psi(t),
            \nonumber \\
\dot G_\epsilon & = &
       -\frac{\partial {\cal C}}{\partial g}(L,G,\ell + \omega t,g)\phi(t) -
                \frac{\partial {\cal S}}{\partial g}(L,G,\ell + \omega t,g)
                    \psi(t),    \nonumber \\
\dot \ell_\epsilon & = &
     -\frac{3k^2}{L^4}L_\epsilon +
       \frac{\partial {\cal C}}{\partial L}(L,G,\ell + \omega t,g)\phi(t)
        +\frac{\partial {\cal S}}{\partial L}(L,G,\ell + \omega t,g)\psi(t),
             \nonumber \\
\dot g_\epsilon & = &  \frac{\partial {\cal C}}{\partial G}
    (L,G,\ell + \omega t,g)
     \phi(t)+\frac{\partial {\cal S}}{\partial G}(L,G,\ell + \omega t,g)\psi(t)
        \nonumber
\end{eqnarray}
with zero initial values;  then, the solution evaluated at $t=m(2\pi/\Omega)$ is
projected to the complement of the range of the infinitesimal
displacement.  It follows from a detailed analysis that one can set the initial
value of time equal to zero, as we have done, with no loss in generality.
The bifurcation function is thus given by
\[
{\cal B}(G,\ell,g)=
(B^L(G,\ell,g),B^G(G,\ell,g),B^{g}(G,\ell,g)),
\]
where
\begin{equation}\label{bifurcationF}
B^L(G,\ell,g) := -\frac{\partial {\cal I}}{\partial \ell}\:, \quad
B^G(G,\ell,g) := -\frac{\partial {\cal I}}{\partial g}\:, \quad
B^g(G,\ell,g) := \frac{\partial {\cal I}}{\partial G}\:,
\end{equation}
and
\[
{\cal I} := \int_0^{2\pi m/\Omega}\left[
          {\cal C}(L,G,\ell+\omega t,g) \phi(t)
+{\cal S}(L,G,\ell+\omega t,g)\psi(t) \right] \,dt.
\]
To evaluate ${\cal I}$, we use the resonance relation $n\Omega = m\omega$ to
change the variable of integration from $t$ to 
$\widehat{\sigma} = \Omega t/m + \ell/n$, 
then observe that the integrand of ${\cal I}$ is periodic with period
$2\pi$ and substitute the Fourier series expansion for  
${\cal C}$ and ${\cal S}$.  
After performing these steps, we obtain the following expression for 
${\cal I}$ in case $n=1$
\begin{eqnarray}\label{calI}
{\cal I} & = & \pi m a^2\Omega 
  \{ U_{2m}(e)\cos{(2g+2m\ell)}
    +\frac{\delta}{2}[(\alpha-\beta)U_{m}(e)\cos{(2g+m\ell)} \nonumber \\
& &\quad +(\alpha+\beta)U_{3m}(e)\cos{(2g+3m\ell)}]\},
\end{eqnarray}
where 
\[U_\nu(e)=A_\nu(e)+B_\nu(e).\]
If $n=2$ and $m$ is odd, then
\[{\cal I} = \pi m a^2 \Omega U_{m}(e)\cos{(2g+m\ell)};\]
while for $n=3$, if $m$ is prime relative to $3$, then
\[
{\cal I} = \frac{1}{2}\pi m a^2 \Omega \delta (\alpha+ \beta)
     U_{m}(e)\cos{(2g+m\ell)}.
\]
It turns out that ${\cal I} = 0$ for $n > 3$, as expected.

We show in Ref.\ \onlinecite{cmr} that the bifurcation function does
not have a simple zero for an incident monochromatic gravitational
wave of definite helicity. The role of the secondary components of the wave
is to resolve this issue for the bifurcation problem under consideration here.
This is indeed the case for $n=1$ as demonstrated in the next section.
However, for $n=2$ and $n=3$ simple zeros do not exist, and the consideration
of these cases would require the calculation of the solutions of
higher order variational initial value problems. On the other hand, higher
order perturbing functions of order $\epsilon^2$, etc., are
neglected in the formulation of equation~(\ref{BasicEq1}), which
is the starting point of our analysis. It follows from this remark
that the treatment of the cases $n=2$ and $n=3$ is beyond the scope of this
work. Thus, we assume $n=1$ in the remainder of this section.

Substituting (\ref{calI}) into (\ref{bifurcationF}), we obtain 
the following explicit form for the bifurcation function ($n=1$)
\begin{eqnarray}\label{bifurcationSet}
B^L(G,\ell,g) & = & 2\pi m^2a^2\Omega \{ U_{2m}(e)
     \sin{(2g+2m\ell)}  \nonumber \\
&  & \hspace*{.25in}  + \frac{\delta}{4}[(\alpha-\beta)U_{m}(e)\sin{(2g+m\ell)}
    \nonumber \\
& & \hspace*{.25in} + 3(\alpha+\beta)U_{3m}(e) \sin{(2g+3m\ell)}]
\}, \nonumber \\
B^G(G,\ell,g) & = & 2\pi ma^2\Omega \{ U_{2m}(e)\sin{(2g+2m\ell)} \nonumber \\
&  & \hspace*{.25in}  + \frac{\delta}{2}[(\alpha-\beta)U_{m}(e)\sin{(2g+m\ell)}
    \nonumber \\
& & \hspace*{.25in} + (\alpha+\beta)U_{3m}(e)\sin{(2g+3m\ell)}] \},
\nonumber \\
B^g(G,\ell,g) & = & -\pi ma^2 \Omega(\frac{G}{eL^2})\{U_{2m}'(e)\cos{(2g+2m\ell)}
\nonumber
\\ & &  +\frac{\delta}{2}[(\alpha-\beta)U_{m}'(e)\cos{(2g+m\ell)}
   \nonumber \\
& &    +(\alpha+\beta)U_{3m}'(e) \cos{(2g+3m\ell)}]\},
\end{eqnarray}
where in the expression for $B^g$ we have used the fact that the eccentricity
$e$ and the Delaunay element $G$ are related by $G = \pm L \sqrt{1-e^2}$.   We
therefore make a change of variable from $G$ to $e$  and   observe that
the zeros and their multiplicities for the bifurcation function ${\cal B}$ are
identical to those of the function
\[
{\cal F}(e,\ell,g)=
(F^L(e,\ell,g),F^G(e,\ell,g),F^{g}(e,\ell,g)),
\]
where
\begin{eqnarray}\label{bifurcation1}
F^L(e,\ell,g) & = &  U_{2m}(e)\sin{(2g+2m\ell)} \nonumber \\
&  &  \hspace*{.25in} 
   + \frac{\delta}{4}[(\alpha-\beta)U_{m}(e)\sin{(2g+m\ell)} \nonumber \\
&   & \hspace*{.3in}  + 3 (\alpha+\beta)U_{3m}(e) \sin{(2g+3m\ell)}],\nonumber \\
F^G(e,\ell,g) & = & U_{2m}(e) \sin{(2g+2m\ell)} \nonumber \\
&  &  \hspace*{.25in} 
  + \frac{\delta}{2}[(\alpha-\beta)U_{m}(e)\sin{(2g+m\ell)} \nonumber \\
&   & \hspace*{.3in}  +  (\alpha+\beta)U_{3m}(e)\sin{(2g+3m\ell)}],\nonumber \\
F^g(e,\ell,g) & = & U_{2m}'(e) \cos{(2g+2m\ell)} \nonumber
\\ & & \hspace*{.25in}
         +\frac{\delta}{2}[(\alpha-\beta)U_{m}'(e)\cos{(2g+m\ell)} \nonumber \\
& &    \hspace*{.3in} +(\alpha+\beta)U_{3m}'(e)\cos{(2g+3m\ell)}].
\end{eqnarray}

To apply Theorem~\ref{conthm},
we must determine the simple  zeros of the
bifurcation function (\ref{bifurcationSet}) or, equivalently,
(\ref{bifurcation1}).
We will show that simple zeros exist by a perturbation argument which is
presented in the next section.

\section{Zeros of a Degenerate Bifurcation Function}
\lb{zdsec}
The bifurcation function (\ref {bifurcation1}) has the following
abstract form
\begin{equation}\label{Delta}
\Delta(\mu,\widehat{\epsilon}) =
  \tau(\mu) + \widehat{\epsilon} \varphi(\mu) = 0, \quad \mu \in {\bbR}^3,
\end{equation}
where the functions $\Delta: {\bbR}^3\times{\bbR}\mapsto{\bbR}^3$ and $\tau,
\varphi: {\bbR}^3 \mapsto {\bbR}^3$ are given in
components by
\[
\tau(\mu) = \left(\begin{array}{c}
\tau_1(\mu) \\ \tau_2(\mu) \\ \tau_3(\mu)
\end{array}\right), \qquad
\varphi(\mu) =  \left(\begin{array}{c}
\varphi_1(\mu) \\ \varphi_2(\mu) \\ \varphi_3(\mu)
\end{array}\right),
\]
and where the first two components of $\tau$ are equal.  In fact, we define
$\widehat{\tau} = \tau_1 = \tau_2 $.
\begin{remark}  The fact that the bifurcation function has the abstract form
(\ref{Delta}) is not accidental.  Every normally incident gravitational plane
wave whose Fourier representation has as its dominant term a purely circularly
polarized  wave will result in a bifurcation function of the form (\ref{Delta})
at resonance.
\end{remark}
We will determine the zero set for $\Delta$ in case
$\widehat{\epsilon}$ is sufficiently small.
\begin{lem}\label{zerosetlem}
Suppose $\Delta$ is defined  by (\ref{Delta}).
If $\eta\in {\bbR}^3$ is such that $\tau(\eta)=0$
while the vectors $\grad \widehat{\tau}(\eta)$ and
$\grad \tau_3(\eta)$ are linearly independent, then there is a curve
$s\mapsto \Gamma(s)$ in ${\bbR}^3$ such that $\Gamma(0)=\eta$,
$\Gamma'(0)\ne 0$, and
$\Delta(\Gamma(s),0)\equiv 0$. If such a curve exists and $s=0$ is a simple zero
of the real-valued function
$s\mapsto \varphi_2(\Gamma(s))-\varphi_1(\Gamma(s))$,  then there is a
curve $\widehat{\epsilon}\mapsto \Upsilon(\widehat{\epsilon})$ in ${\bbR}^3$
such that $\Upsilon(0)=\eta$ and
$\Delta(\Upsilon(\widehat{\epsilon}),\widehat{\epsilon})\equiv 0$.
Moreover, for each sufficiently small $\widehat{\epsilon}\ne 0$, the point
$\Upsilon(\widehat{\epsilon})\in {\bbR}^3$ is a simple zero of the function
$\mu\mapsto \Delta(\mu,\widehat{\epsilon})$.
\end{lem}
{\em Proof}.
The linearly independent vectors $\grad \widehat{\tau}(\eta)$ and $\grad
\tau_3(\eta)$ span a plane $Q$ in ${\bbR}^3$. Let $S$ denote one of the
two possible rotation operators in space that preserves $Q$ and rotates each
vector in $Q$ through $\pi/2$ radians. 
We note that $S\grad \widehat{\tau}(\eta)$
and $S\grad \tau_3(\eta)$ are linearly independent vectors in $Q$.
Also, we let $v$ denote a vector in space such that the
set $\{S\grad \widehat{\tau}(\eta),S\grad \tau_3(\eta), v\}$ is linearly
independent. This requirement is satisfied, for instance, if we
choose $v$ to be equal to the cross product of
$S\grad \widehat{\tau}(\eta)$ and $S\grad \tau_3(\eta)$. 

The function $w:{\bbR}^3\mapsto{\bbR}^3$ defined by
\[
(\mu_1,\mu_2,\mu_3)\mapsto \mu_1S\grad \tau_3(\eta)+\mu_2S\grad
\widehat{\tau}(\eta)+\mu_3 v +\eta
\]
is invertible. We use it to define $\Theta:{\bbR}^3\to{\bbR}^2$ given by
\[
\Theta(\mu_1,\mu_2,\mu_3)=(\widehat{\tau}
(w(\mu_1,\mu_2,\mu_3)),\tau_3(w(\mu_1,\mu_2,\mu_3)).
\]
It follows that $\Theta(0)=0$. Moreover, 
the derivative of the transformation
\[
(\mu_1,\mu_2)\mapsto (\widehat{\tau}(w(\mu_1,\mu_2,0)),\tau_3(w(\mu_1,\mu_2,0))
\]
at $(\mu_1,\mu_2)=(0,0)$ is given by the matrix
\begin{equation}\label{dermat}
\left(
\begin{array}{cc}
\grad \widehat{\tau}(\eta)\cdot S\grad\tau_3(\eta) & 0 \\
0&\grad \tau_3(\eta)\cdot S\grad \widehat{\tau}(\eta)
\end{array}
\right)
\end{equation}
that represents the partial derivative of $\Theta$ with respect to its
first two arguments at the origin.

Using the linear independence of $\grad \widehat{\tau}(\eta)$ and $\grad
\tau_3(\eta)$, it is easy to see that the diagonal elements of the matrix
(\ref{dermat}) are both nonzero. Thus, by the Implicit Function Theorem, there
is a unique curve $s\mapsto \sigma(s)$ in the $(\mu_1, \mu_2)$-plane such that
$\sigma(0)=(0,0)$
and $\Theta(\sigma(s),s)\equiv 0$. The curve $\Gamma(s):=w(\sigma(s),s)$
is such that $\Gamma(0)=\eta$ and $\Delta(\Gamma(s),0)\equiv 0$.
Moreover,
\[
\Gamma'(0)=Dw(0)\left(
\begin{array}{c}
\sigma'(0) \\
1
\end{array}
\right).
\]
Since $Dw(0)$ is invertible and the vector $(\sigma'(0),1)\ne 0$,
we have that $\Gamma'(0)\ne 0$. This proves the first assertion
of the theorem.

Under the assumption that $s=0$ is a simple zero of the function
$s\mapsto \varphi_2(\Gamma(s))-\varphi_1(\Gamma(s))$, we will use the
Lyapunov-Schmidt reduction procedure to show that the corresponding point
$\eta=\Gamma(0)$ in
${\bbR}^3$ is continuable as a zero of $\Delta$.
As $\Gamma'(0)\ne 0$,
the image $\widehat{\cal Z}$ of $\Gamma$ is locally
a one dimensional submanifold of ${\bbR}^3$.
Moreover, $\widehat{\cal Z}$ is normally nondegenerate in the sense that
at each point $\widehat{z}\in \widehat{\cal Z}$ sufficiently close to $\eta$, the kernel
$\widehat{K}$ of
$D\tau(\widehat{z})$ is exactly the one dimensional tangent space of $\widehat{\cal Z}$.
In fact, since this tangent space is clearly in $\widehat{K}$, it suffices to show
 that
$\widehat{K}$ is one dimensional. The derivative of $\tau$ is expressed in the
standard vector notation by
\[
D\tau(\widehat{z})=\left(\begin{array}{c}
\grad \widehat{\tau}(\widehat{z}) \\
\grad \widehat{\tau}(\widehat{z}) \\
\grad \tau_3(\widehat{z}) \\
\end{array} \right).
\]
Since the vectors $\grad \widehat{\tau}(\widehat{z})$ and $\grad \tau_3(\widehat{z})$ are
linearly independent at $\widehat{z}=\eta$, they will, by continuity,  remain
linearly independent in an open neighborhood of $\eta$.
Thus, their span is  two dimensional at each point of this neighborhood.
In particular,  for $\widehat{z}$ in this neighborhood, it is clear
(by matrix multiplication)
that $\grad \widehat{\tau}(\widehat{z})$ and $\grad \tau_3(\widehat{z})$ are not both in
$\widehat{K}$. This proves that $\widehat{K}$ is one dimensional.
It follows from  
the same Lyapunov-Schmidt reduction procedure that is used to prove Theorem
\ref{conthm} (see Ref.\ \onlinecite{ccc}) that if $\Pi(\widehat{z})$ denotes a
projection to the complement
$\widehat{\cal S}$ of the range of 
$D\tau(\widehat{z})$, then simple zeros of the
map from $\widehat{\cal Z}$ to $\bbR$ given by
$\widehat{z}\mapsto \Pi(\widehat{z})\Delta_{\hat{\epsilon}}(\widehat{z},0)$
are continuable. Thus, if $\eta$ is such a simple zero, there is a curve
$\widehat{\epsilon}\mapsto \Upsilon(\widehat{\epsilon})$ 
such that $\Upsilon(0)=\eta$
and $\Delta(\Upsilon(\widehat{\epsilon}),\widehat{\epsilon})\equiv 0$ as 
required in the
lemma.

To construct such a projection into $\widehat{\cal S}$, it can be shown
that, in fact,
the range of $D\tau(\widehat{z})$ 
is spanned by the transpositions of the vectors
$(1,1, 0)$ and $(0,0,1)$. To see this,
it is sufficient to note that 
$D\tau(\widehat{z})S \grad \widehat{\tau}(\widehat{z})$ is the transpose of
a scalar multiple of $(0,0,1)$ while
$D\tau(\widehat{z})S\grad\tau_3(\widehat{z})$ is a scalar multiple of
the transpose of $(1,1,0)$. 
The transpose of the vector $(0,1,0)$ clearly spans a complement to
the  range that we will denote by $\widehat{\cal S}$. Thus, if
$(\rho_1,\rho_2,\rho_3)\in {\bbR}^3$, 
the projection $\Pi$ is easily computed and is given by
$\Pi(\rho_1,\rho_2,\rho_3)= \rho_2 - \rho_1$.
Note that $\Pi$ does not depend on the base point 
$\widehat{z}\in\widehat{\calZ}$.

Using the projection $\Pi$ and the definition of $\Delta$,
the Lyapunov-Schmidt reduced function
$\widehat{z}\mapsto \Pi(\widehat{z})\Delta_{\hat{\epsilon}}(\widehat{z},0)$
is given by 
$\widehat{z}\mapsto \varphi_2(\widehat{z})-\varphi_1(\widehat{z})$ for
$\widehat{z}\in \widehat{\cal Z}$. 
Thus, if, in the coordinates of $\widehat{\cal Z}$, $s=0$ is a
simple zero of $s\mapsto \varphi_2(\Gamma(s))-\varphi_1(\Gamma(s))$, then
$\eta=\Gamma(0)$ is a continuable zero of $\Delta$ given by a curve $\Upsilon$,
as required in the statement of the lemma.

It remains to show that 
$\Upsilon(\widehat{\epsilon})$, for sufficiently small
$\widehat{\epsilon}\ne 0$,
is in fact a {\em simple}
zero of the function $\mu \mapsto \Delta(\mu,\widehat{\epsilon})$.
To this end, it suffices to show that the matrix
$D\Delta(\Upsilon(\widehat{\epsilon}),\widehat{\epsilon})$
does not have a zero eigenvalue. But, under the hypotheses of the lemma,
the matrix $D\Delta(\eta,0)$ has a one dimensional kernel, namely the tangent
space of $\widehat{\cal Z}$. 
This means that  $D\Delta(\eta,0)$ has exactly one zero eigenvalue. \
By the continuity of eigenvalues of matrices, there is a smooth
family of eigenvalues 
$\lambda(\widehat{\epsilon})$ such that $\lambda(0)=0$ and
a corresponding smooth family of eigenvectors $V(\widehat{\epsilon})$ such that
$V(0)$ is a nonzero vector tangent to $\widehat{\cal Z}$ with
\[ 
D\Delta(\Upsilon(\widehat{\epsilon}),\widehat{\epsilon}) 
  V(\widehat{\epsilon}) =
 \lambda(\widehat{\epsilon})V(\widehat{\epsilon}).
\]

It suffices to show that if $\widehat{\epsilon}$ is positive and
sufficiently small, then $\lambda(\widehat{\epsilon})\ne 0$. 
By continuity, the remaining two eigenvalues of the matrix will be
nonzero as well. The desired result follows as soon as we show that
the derivative $\lambda'(0)\ne 0$. For this we have
\[
D^2\Delta(\eta,0)(\Upsilon'(0),V(0))+D\Delta_{\hat{\epsilon}}(\eta,0)V(0)
+D\Delta(\eta,0) V'(0)=\lambda'(0)V(0),
\]
where $D$ denotes differentiation with respect
to the space variable $\mu \in {\bbR}^3$.
After projection by $\Pi$ into the complement of the 
range of $D\Delta(\eta,0)$,
\[
\Pi D^2\Delta(\eta,0)(\Upsilon'(0),V(0))
   +\Pi D \Delta_{\hat{\epsilon}}(\eta,0) V(0)
    = \lambda'(0) \Pi V(0).
\]

We claim that $\Pi D^2\Delta(\eta,0)(\Upsilon'(0),V(0))=0$. Once this claim is
proved,
the fact that $\lambda'(0)\ne 0$  would follow provided
$\Pi D \Delta_{\hat{\epsilon}}(\eta,0) V(0)\ne 0$.
The key point to note is that the projection operator, $\Pi$, is independent of
the base point, i.e. the range of the projection operator is spanned by a
constant basis vector independent of the base point.
Thus since $\Delta_{\hat{\epsilon}}(\Gamma(s),0)=\varphi(\Gamma(s))$,
we have
\[
\left.\frac{d}{ds}\Pi \varphi(\Gamma(s))\right |_{s=0}=\Pi D
\varphi(\eta)\Gamma'(0) =\Pi D\Delta_{\hat{\epsilon}}(\eta,0)\Gamma'(0).
\]
Since $\widehat{\cal Z}$ is one dimensional, 
$\Gamma'(0)$ is just a scalar multiple
of $V(0)$ and therefore if $\Pi D \varphi(\eta)\Gamma'(0)\ne 0$, then
$\Pi D\Delta_{\hat{\epsilon}}(\eta,0)V(0)\ne 0$. By the definition of the
components
of $\varphi$ and of the projection  $\Pi$, we have
\[
\left.\frac{d}{ds}\Pi \varphi(\Gamma(s))\right |_{s=0}=
\left.\frac{d}{ds}[\varphi_2(\Gamma(s))-\varphi_1(\Gamma(s))]\right |_{s=0}.
\]
Thus, if $s=0$ is a simple zero of
$s\mapsto \varphi_2(\Gamma(s))-\varphi_1(\Gamma(s))$, then
\[\Pi D\Delta_{\hat{\epsilon}}(\eta,0)V(0)\ne 0,\] as required.

To verify the claim, note that since $\Pi$ projects to the complement of the
range of  $D\Delta(\eta,0)$ and since $\Pi$ does not depend on the
base point on $\widehat{\calZ}$, we have
$\Pi D\Delta(\Gamma(\vartheta),0)\Upsilon'(\vartheta)\equiv 0$ for the
real variable $\vartheta$ in the common domain of $\Gamma$ and $\Upsilon$.
This implies that
\begin{eqnarray}
0 & = & \left.\frac{d}{d\vartheta}
  [\Pi D\Delta(\Gamma(\vartheta),0)\Upsilon'(\vartheta)]\right |_{\vartheta=0}
                \nonumber \\
  & = & \Pi D^2\Delta(\eta,0)(\Gamma'(0),\Upsilon'(0))+\Pi
               D\Delta(\eta,0)\Upsilon''(0) \nonumber \\
  & = & \widehat{c}\Pi D^2\Delta(\eta,0)(V(0),\Upsilon'(0)), \nonumber
\end{eqnarray}
where $\widehat{c}V(0)=\Gamma'(0)$. 
Since $\widehat{c}\ne 0$, the lemma is proved.
\section{Periodic Orbits}
\lb{secpo}
We now use Lemma~\ref{zerosetlem} to prove that
some of the zeros of the bifurcation function
(\ref{bifurcation1}) are simple.  We note that once this
result is established, it will follow from Theorem~\ref{conthm}
that the corresponding bounded
orbits of the Keplerian two-body system are continuable under perturbation by
periodic gravitational waves.

If $\delta = 0$, then the zeros of (\ref{bifurcation1})  are the union of the
following one dimensional sets
\begin{eqnarray}
{\calZ}_1^+ & = & \{ (e,\ell,g) : 2g+2m\ell = 0,\quad
  U_{2m}'(e) =  0 \},\nonumber \\
{\calZ}_1^- & = & \{ (e,\ell,g) : 2g+2m\ell = \pi,\quad
  U_{2m}'(e) =  0 \},\nonumber \\
{\calZ}_2^+ & = & \{ (e,\ell,g) : 2g+2m\ell = \frac{\pi}{2}, \quad
      U_{2m}(e) = 0 \}, \nonumber \\
 {\calZ}_2^- &=& \{(e,\ell,g) : 2g+2m\ell = \frac{3\pi}{2}, \quad
      U_{2m}(e) = 0 \}.\lb{zeroset}
\end{eqnarray}
We will show that the zeros in the sets ${\calZ}_1^{\pm}$ and ${\calZ}_2^{\pm}$
continue to simple zeros of the bifurcation
function for sufficiently small $\delta \neq 0$.  To conform with
Lemma~\ref{zerosetlem}, we use  (\ref{bifurcation1}) to identify the components
of $\Delta$ as they appear in the lemma as follows:
\begin{eqnarray}
\widehat{\tau}(e,\ell,g) & = &  U_{2m}(e) \sin{(2g+2m\ell)}, \nonumber \\
\tau_3 (e,\ell,g) & = & U_{2m}'(e)\cos{(2g+2m\ell)}, \nonumber \\
\varphi_1(e,\ell,g) & = & 
   \frac{1}{4} \left[ (\alpha - \beta)U_{m}(e)\sin{(2g+m\ell)} \right. \nonumber \\
&  & \hspace*{.5in} \left.
+3(\alpha + \beta)U_{3m}(e)\sin{(2g+3m\ell)} \right], \nonumber \\
\varphi_2(e,\ell,g) & = & 
  \frac{1}{2} \left[ (\alpha - \beta)U_{m}(e)\sin{(2g+m\ell)} \right. \nonumber \\ 
&  & \hspace*{.5in} \left. +
(\alpha + \beta)U_{3m}(e)\sin{(2g+3m\ell)} \right], \nonumber \\
\varphi_3(e,\ell,g) & = & 
  \frac{1}{2}\left[ (\alpha - \beta) U_{m}'(e)\cos{(2g+m\ell)}\right. \nonumber \\
&  &  \hspace*{.5in} \left.  + (\alpha + \beta) U_{3m}'(e)\cos{(2g+3m\ell)}\right].
\nonumber
\end{eqnarray}
Also, we note that $\delta$ in (\ref{bifurcation1}) plays the role of
$\hat{\epsilon}$ in the lemma.

There are four cases to consider corresponding to the zero sets (\ref{zeroset})
of the unperturbed bifurcation function (\ref{bifurcationSet}).
We will consider the zero set ${\cal Z}_1^+$
for illustrative purposes, as the computational procedure
is identical in all four cases.   The set ${\calZ}_1^+$ is just a line in
${\bbR}^3$, so here the curve $\Gamma$ in the lemma can be taken to be a
parametrization of this line starting at an appropriate point.
By a direct calculation, the function
$\varphi_2 - \varphi_1$ on ${\cal Z}_1^+$ is given by
\[
\frac{1}{4}\left[(\alpha - \beta)U_{m}(e) + (\alpha+\beta)U_{3m}(e)\right]\sin{g}.
\]
The zeros of this function along ${\calZ}_1^+$ are clearly simple provided
the coefficient of $\sin g$ does not vanish at the value of the eccentricity
determined by membership in ${\calZ}_1^+$.  The main result of
this section is the following theorem.
\begin{thm}\label{zerosetsun}
Consider the system~(\ref{EqofMotion}) and
suppose that $(L,G,\ell,g)$ is on an unperturbed $(m:n)$ 
resonant periodic solution with $n=1$, that is,  $L^3=mk^2/\Omega$.
If  $(e,\ell,g)$ is in ${\calZ}_1^+$ or ${\calZ}_1^-$, respectively,
$(e,\ell,g)$ is in ${\calZ}_2^+$ or ${\calZ}_2^-$, where 
$e$ is the eccentricity of the corresponding Keplerian ellipse
($e^2=1-G^2/L^2$) and if
\[
(\alpha - \beta)U_{m}(e)+(\alpha + \beta)U_{3m}(e)\ne 0,
\]
respectively,
\[
(\alpha - \beta)U_{m}(e)-(\alpha + \beta)U_{3m}(e)\ne 0,
\]
then the Keplerian ellipse continues  to
a periodic orbit under the perturbation. 
\end{thm}
It remains to show that the sets $\{{\calZ}_i^{\pm}\}$, $i=1,2$, are not
empty. This fact will follow as soon as we show that both of the functions
$U_{2m}$ and $U_{2m}'$ have zeros on the interval $0<e<1$.

Recall that  
$U_{2m}(e)=A_{2m}(e)+B_{2m}(e)$ 
and note that both $A_{2m}$ and $B_{2m}$ have
removable singularities at $e = 0$.  
Moreover, both have Taylor series at $e = 0$ with leading terms given by
\[2(2m-1)\frac{m^{2m-2}}{(2m)!}e^{2(m-1)}.\] 
In particular, if $m=1$, then
$\lim_{e \rightarrow 0^+} U_{2}(e) = 2$  and for $m > 1$, the limit is
zero, but $U_{2m}(e) > 0$ for sufficiently small eccentricity.
Also, $\lim_{e \rightarrow 1^-} U_{2m} = -J_{2m}(2m)/m^2$.  
By a standard property of the Bessel functions,
$J_{\nu}(\nu) > 0$; hence, $U_{2m}(1) < 0$.  This proves $U_{2m}$ has at
least one zero on the interval $0 < e < 1$. 
Numerical calculations suggest that this zero is unique.

A simple argument can be given to prove the existence of a zero in the case
$m=1$ for the function $U_{2m}'$.
In this case, the series expansion of $U_2'$ at $e = 0$ is $-10 e + O(e^2)$.
Thus,
$U_2'(e)<0$ for a small but positive eccentricity.  
The limit as $e \rightarrow 1^-$ is the same as
\[ \lim_{e \rightarrow 1^-} \frac{2}{\sqrt{1-e^2}} J_{2}'(2).\]
By standard properties of the Bessel functions, $J_2(0)=0$ and $J_2 > 0$ on the
interval $(0, j'_{2,1})$, where $j'_{2,1}$ is the first zero of
$J_{2}'$.  Hence, $J_{2}' > 0$ on $(0, j'_{2,1})$.  But, we also
have $2 < j'_{2,1}$.  Thus, $J_{2}'(2) > 0$ and $U_{2}'(e)
\rightarrow \infty$ as $e \rightarrow 1^{-}$.  
This proves that $U_{2}'(e)$
has at least one zero on the interval $0<e<1$; 
moreover, numerical computations suggest that
this zero is unique.

For $m\ge 2$, we will outline a proof that shows the function $U_{2m}'$ has
at least two zeros on the interval $0<e<1$.  
An asymptotic analysis
shows that the function $U_{2m}'$ 
is positive near the end points of the interval. 
The proof is completed by showing that
$U_{2m}'$ has a negative value within the interval.
To this end, note that the function $U_{2m}'$ has the form
\[ U_{2m}'(e)=\gamma_1(m,e)J_{2m}(2me)+\gamma_2(m,e)J_{2m}'(2me),\]
where $\gamma_1$ and $\gamma_2$ are computed using~(\ref{ABessel}).
Moreover, by standard properties of the Bessel functions,
both of the functions $e\mapsto J_{2m}(2me)$ and $e\mapsto J_{2m}'(2me)$
are positive for $0<e<1$. 
Let us define $\widehat{e}:=(1-1/(4m^2))^{1/2}$,
and observe that $0<\widehat{e}<1$. 
An simple computation shows that
\[
\gamma_1(m,\widehat{e})<0,\qquad \gamma_2(m,\widehat{e})<0.
\]
Hence, we have  $U_{2m}'(\widehat{e})<0$, as desired. 
Numerical computations suggest that the function $U_{2m}'$ 
has exactly two zeros on the interval $0<e<1$.

It follows from these results that the resonant interaction of the incident
multi-frequency gravitational wave of definite helicity with a
Keplerian binary can result in orbits that are periodic with period
$2\pi/\Omega$, where $\Omega=m\omega$, $m=1,2,3,\ldots$. It is conceivable
that other periodic orbits may exist; however, our method can identify
only those periodic orbits that are continuations of resonant Keplerian
orbits of the unperturbed system.
\section{Rotating Frame}
\lb{rfsec}
In a manner similar to that used by Hill in his treatment of the lunar
theory \cite{Kov,poincare,Stern}, we can view the dynamical system described by
(\ref{Ham1}) in Cartesian coordinates rotating at half the principal frequency
of the incident gravitational wave $(\Omega_{\star}=2\Omega)$.  These
coordinates---that we again represent by ($x,y$)---rotate with frequency
$\frac{1}{2}\Omega_{\star} = \Omega$ with respect to inertial coordinates;
therefore, the equations of motion in these coordinates are given by
\begin{eqnarray}\label{HamRot}
\frac{d^2x}{dt^2} - 2\Omega \frac{dy}{dt} - \Omega^2 x + \frac{kx}{r^3}+
2\epsilon \Omega^2\left[\left(1+\delta\alpha\cos{\Omega t}\right)x\right.
\hspace*{.5in} & & \nonumber \\
\left. +\left(\delta\beta\sin{\Omega t}\right)y\right] & = & 0, \nonumber \\
\frac{d^2y}{dt^2} + 2\Omega \frac{dx}{dt} - \Omega^2 y + \frac{ky}{r^3} +
2\epsilon \Omega^2\left[\left(\delta\beta\sin{\Omega t}\right)x\right.
\hspace*{.5in} & & \nonumber \\
\left. -\left(1+\delta\alpha\cos{\Omega t}\right)y\right] & = & 0.
\end{eqnarray}
 Let $X := \dot{x} - \Omega y$ and $Y := \dot y + \Omega x$ be the canonical
momenta conjugate to $x$ and $y$, respectively.  Then (\ref{HamRot}) is
equivalent to a Hamiltonian system with Hamiltonian
\begin{eqnarray}\label{HamCart}
{\cal H}_R & = & \frac{1}{2}(X^2+Y^2) + \Omega(yX-xY) - \frac{k}{r} \nonumber \\
& & + \epsilon \Omega^2[(1+\delta\alpha\cos{\Omega t})(x^2-y^2)+
     2(\delta\beta\sin{\Omega t}) xy].
\end{eqnarray}
By identifying the momenta in polar rotating coordinates as $p_r = (xX+yY)/r$
and $p_{\theta} =xY-yX$,  (\ref{HamCart}) in polar coordinates is given by
\begin{eqnarray}\label{HamPolar}
{\cal H}_R = \frac{1}{2}\left(p_r^2+\frac{p_{\theta}^2}{r^2}\right) -
  \frac{k}{r} - \Omega p_{\theta} + \epsilon \Omega^2r^2 [ \cos{2\theta}
      \hspace*{.5in}  & & \nonumber \\ 
+ \delta (\alpha\cos{2 \theta} \cos{\Omega t}
       + \beta \sin{2 \theta}\sin{\Omega t})], & &
\end{eqnarray}
which in the corresponding Delaunay elements becomes
\begin{eqnarray}\label{HamDelaunay}
{\cal H}_R  = -\frac{k^2}{2L^2}-\Omega G +\epsilon\Omega^2 \{ {\cal
C}(L,G,\ell,g) + \delta [ \alpha{\cal C}(L,G,\ell,g)\cos{\Omega t}   & &
\nonumber \\
+\beta {\cal S}(L,G,\ell,g)\sin{\Omega t}] \}. & &
\end{eqnarray}

The Hamiltonian system given by (\ref{HamDelaunay}) has periodic orbits with
period $2\pi/\Omega$, where $\Omega = m \omega$, $m = 1, 2, \cdots$; this
assertion can be demonstrated using results that have already been obtained in
this paper.  In fact,
using the resonance assumption, periodic orbits exist in the inertial frame with
period $2\pi/\Omega$ as shown in Section~\ref{secpo}.  
An orbit in the inertial frame,
$t\mapsto(x_I(t), y_I(t))$, is represented in the rotating frame by
\begin{eqnarray*}
x_R(t) & = & x_I(t) \cos{\Omega t} + y_I(t) \sin{\Omega t}, \\
y_R(t) & = & -x_I(t) \sin{\Omega t} + y_I(t) \cos{\Omega t}.
\end{eqnarray*}
It follows from these relations that if $x_I(t)$ and $y_I(t)$ are
periodic with period $2\pi/\Omega$, then so are $x_R(t)$ and $y_R(t)$.
Therefore, the orbits that are periodic in the inertial frame with period
$2\pi/\Omega$ are observed to be periodic with period $2\pi/\Omega$ in the
rotating frame.  It is also possible to arrive at this conclusion by direct
application of the methods developed in the previous sections.

To understand the physical structure of Hamiltonian (\ref{HamDelaunay}),
imagine right circularly polarized gravitational radiation of frequency
$2\Omega$ that is normally incident on the orbital plane.  It follows from our
previous work \cite{cmr}, as well as equation (\ref{HamRot}), that in a
reference frame rotating with frequency $\Omega$ the wave stands still.  That
is, observers at rest in the rotating frame do not perceive the variability
associated with a wave so that in the absence of secondaries ($\delta = 0$) the
dynamical system (\ref{HamRot}) is autonomous.  That an observer---by merely
rotating about the propagation axis of the circularly polarized wave---could
make the wave stand completely still would be a remarkable physical effect and
deserves further discussion.

It is a fundamental consequence of Lorentz invariance 
that all basic radiation
fields travel with speed $c$ with respect to all {\em inertial} observers.  
This
may be illustrated by an example:  Let an inertial observer move with speed
$v_0$ along the propagation axis of a monochromatic plane gravitational wave of
frequency $\Omega_{\star}$.  The frequency and the wave vector of the radiation
as perceived by the moving observer are smaller than those measured by static
inertial observers by a common Doppler factor of
\[\left(\frac{c-v_0}{c+v_0}\right)^{\frac{1}{2}}.\]
Mathematically, as $v_0 \rightarrow c$ this ratio goes to zero and hence the
frequency and wave vector of the radiation vanish so that the wave might appear
to stand still.  This limit is not physically allowed, however.  
No observer can
move at the speed of light, although---theoretically---one can get
arbitrarily close.  Therefore,  the wave can never stand still for an inertial
observer.  It has been shown \cite{mashhoon3} that according to the standard
Einstein theory this is not the case for accelerated observers, i.e. an
accelerated observer can indeed stand still with respect to a gravitational
wave.  The autonomous nature of the system (\ref{HamRot}) for $\delta = 0$
provides an interesting illustration of this fact.  That is, consider an
observer at the center of a system of coordinates rotating with frequency
$\Omega$.  The observer does not move, but the fact that it refers its
observations to the axes that rotate with frequency $\Omega$ with respect to the
inertial axes makes it a noninertial observer.  Radiation of frequency
$\Omega_{\star}$ is incident in the inertial frame along the axis of rotation.
According to the noninertial observer, the frequency of the gravitational wave
is $\Omega_{\star}' = \Omega_{\star} \mp 2 \Omega$, where the upper sign
refers to  right circularly polarized (RCP) gravitational radiation and the
lower sign refers to  left circularly polarized (LCP)  gravitational radiation.
The first (second) case has helicity +2 ($-2$), so that $\Omega_{\star}' =
\Omega_{\star} - {\bf h} \cdot {\bf \Omega}$, where ${\bf h}$ is the helicity
of the gravitational radiation field; this is an example of the general
phenomenon of helicity-rotation coupling.  Now if $\Omega = \frac{1}{2}
\Omega_{\star}$ for RCP gravitational waves or  $\Omega = -\frac{1}{2}
\Omega_{\star}$ for LCP gravitational waves, we find that $\Omega_{\star}'$
vanishes according to the noninertial observer (as well as any other observer at
rest in the rotating frame anywhere along the $z$-axis) and the radiation field
stands still.  It has been shown in Ref.\ \onlinecite{cmr} 
that for a monochromatic
gravitational wave with definite helicity and sufficiently small 
amplitude this observation concerning a rotating observer 
leads to the conclusion that a
Keplerian system in the presence of this radiation can never ionize.  
{\em In principle, this absence of ionization could be considered an 
observable consequence of the physical possibility that a gravitational 
wave could stand completely still}.

Let us now consider the possibility of ionization of the binary system
as a function of the parameter $\delta$.  The transformation to the rotating
frame leaves the orbital radius unchanged; therefore, the ionization problem can
be discussed equally well in the rotating frame.  In fact, the problem becomes
simpler since the principal component of the incident radiation field loses its
time - dependence as in (\ref{HamDelaunay}).  Thus if $\delta = 0$, the KAM
theorem implies that for sufficiently small $\epsilon$ the perturbed trajectory
is bounded since it is trapped between two-dimensional invariant tori in the
three-dimensional energy surfaces.  When $\delta \neq 0$, however, the two
secondary components in the inertial frame both reduce to a perturbation of
frequency $\Omega$ in the rotating frame; that is, the three Fourier components
of the radiation field in the inertial frame are RCP waves with frequencies
$\Omega_{\star} = \Omega, 2 \Omega$ and $3 \Omega$, while the tidal matrix in
the rotating frame has frequencies given by $\Omega_{\star}' = \Omega_{\star} -
2\Omega$, i.e. $-\Omega, 0$ and $\Omega$ for the three components,
respectively.  The perturbation of frequency $\Omega$ in the rotating frame is
expected to lead to Arnold diffusion \cite{arnold,liberman} and hence ionization
of the system.  However, it has not been possible thus far to prove ionization
for the system (\ref{HamDelaunay}); therefore, we resort to numerical work in
the
following section.

\section{Numerical Experiments}
\lb{nesec}
In this section we illustrate---by means of numerical
experiments---the conjecture that gravitational ionization and Arnold
diffusion are closely related.  The interpretation of the numerical results is
simplified if we take the viewpoint of inertial observers and consider the
perturbed motion given by Hamiltonian (\ref{Ham1}).

We have performed several numerical experiments to test the diffusion and
ionization properties of the dynamical system that is represented
by the Hamiltonian~(\ref{Ham1}).   
The physical meaning of
these numerical experiments is essentially the same as
in our previous paper (cf.\  Figure 2 in Ref.\ \onlinecite{cmr}): 
Let us choose two scales for
the measurement of time and length that are arbitrary except that they are
connected here by our choice of $k = 1$.  In these otherwise unspecified units,
we have chosen an unperturbed ellipse of semimajor axis $a = 4/3$ and
eccentricity $e=1/2$ such that $g = -\pi/2$ and the Keplerian frequency is
$\omega = 3\sqrt{3}/8$.  The ellipse is perturbed by the presence of periodic
gravitational radiation and the orbital radius of the osculating ellipse is then
plotted versus time in these units in Figures~1--4.  These figures represent the
results of our numerical experiments in which we have
set the parameters of the external perturbation as follows:
\[
\epsilon=.001,\quad\alpha=2.5,\quad\beta=2,
\]
and have changed $\delta$ in the range $0 \le \delta \le 1$ and
$\Omega$ in the range $1 \le \Omega \le 2$. In each run,
we set the initial conditions to be
\[(p_r,p_\theta,r,\theta)= (.5, 1 , 1, 0),\]
which correspond to the unperturbed ellipse described above.
After integration over each time interval corresponding to
one cycle of the perturbation, i.e. $2\pi/\Omega$, the corresponding
elapsed time $t$ and orbital radius are plotted.  Clearly, for each value of $t$
(abscissa) there is only one value of orbital radius $r$ (ordinate); however,
this is not discernible in some of the figures due to the way in which the plots
have been prepared.  The KAM theorem is illustrated in Figure~1, where the value
of $r$ appears to oscillate between $\simeq a(1-e) = 2/3$ and 
$\simeq a(1+e) = 2$, thus indicating the complete absence of ionization.  
The possibility of
ionization of the system is illustrated in Figures 2 and 3 for $\delta > 0$,
where the initial ellipse is near resonance in the top panel, on resonance in
the middle panel and off resonance in the bottom panel.  
More precisely, the middle panel in
either Figure~2 or Figure~3 
corresponds to an exact third-order resonance since $\Omega/\omega = 3$, 
while the top panel illustrates the response of the ellipse to the
external perturbation near resonance $\Omega/\omega \approx 3.08$; the bottom
panel is off resonance with $\Omega/\omega \approx 1.54$.  
Additional calculations extending the integration time 
for the system depicted in the top panel of Figure~3 
have been performed. These results suggest that a bursting 
behavior occurs in which the near
resonance condition shown at the right end of this panel is followed by chaotic
motion similar to that shown in the middle of this panel that in turn is
followed by a period of near resonance.  This recurrence of chaotic and near
resonance behavior appears to continue for the extended
interval of time studied as illustrated in Figure~4. Thus, this behavior
is consistent with a type of chaotic behavior, called intermittent
chaos, that has been studied for dissipative systems~\cite{liberman}. 
If the behavior suggested by
these simulations is indeed present in the Hamiltonian system~(\ref{Ham1}), 
then our result would be an example of {\em Hamiltonian intermittency}.

Imagine, for the sake of concreteness, a binary system consisting of an
artificial satellite in an eccentric orbit about the Earth. Let the
scales of length and time be $R_0$ and $T_0$, respectively;
then, $R_0^3=kT^2_0$. Thus,  if we take $R_0=10^9\; \cm$ for the problem
under consideration, it turns out that $T_0\approx 1.6\times 10^3\; \sec$.
The gravitational wave in our numerical experiments would then have
a frequency of the order of $\Omega\approx 10^{-3}\; \rad\; \sec^{-1}$, 
corresponding approximately to $1.5\times 10^{-4}\; \Hz$ 
as well as to a wavelength of 
$2\times 10^{14}\; \cm$, and an amplitude of the
order of $\epsilon=10^{-3}$. 
Gravitational waves have not yet been
directly observed; however, in a realistic situation the amplitude of the
wave would be expected to be of the order of $10^{-20}$.

Our numerical experiments are consistent with the expected behavior for a
2$\frac{1}{2}$-degree of freedom Hamiltonian system. 
In fact, by introducing a
fictitious action variable (cf. \S 6 in Ref.\ \onlinecite{cmr}), our system 
is equivalent to an {\em autonomous} Hamiltonian system with 
{\em three} degrees of freedom.
Each orbit of this new system is
constrained to an energy manifold. However,
whereas the two dimensional KAM tori (if they exist for our
choice of parameter values) separate each
three dimensional energy manifold for the corresponding
two-degree of freedom
Hamiltonian system that we obtain with $\delta=0$,
the three dimensional KAM tori, that may exist for the three-degree of 
freedom Hamiltonian system that we obtain with $\delta\ne 0$, do
not separate space within the five dimensional energy manifolds.
Of course, for sufficiently small choices of $\delta$, 
there are orbits of the three-degree of freedom Hamiltonian
system that remain bounded for all time; for example, 
the periodic orbits of Section~\ref{secpo} and the orbits
confined to KAM tori. While the totality of bounded orbits in an
energy manifold may be a set of
positive measure, we expect that every open set of the five
dimensional energy manifold contains an initial condition for a 
trajectory that will diffuse throughout the energy manifold. In fact,
we expect this behavior for all $\delta>0$. 
On the other hand, as $\delta$ decreases toward zero, the time
required to leave the vicinity of a KAM torus  
is expected to grow at an exponential rate.  Figures~1--4
illustrate dynamical behavior that is characteristic of Arnold diffusion
\cite{liberman}.
\section{Appendix}
\lb{appsec}
\appendix
The purpose of this appendix is to point out that the main results of
this paper still hold for a more general incident gravitational wave than that
considered in  (\ref{chimatrix}).

In the general case of incident gravitational radiation on a binary system, the
motion of the system away from the initial orbital plane (i.e. along the
$z$-direction) also needs to be taken into account.  It is necessary to mention
here that {\em initial conditions} are generally ignored in our theoretical
approach, which
relies on the properties of the perturbed system once transients have died away
and a ``steady state" situation has been established.  In this paper, we have
limited our considerations to normally incident waves; in fact, the
transversality of gravitational radiation makes it possible to set $z = 0$ for
normally incident waves.  We wish to note here that essentially the same results
can be obtained for a more general incident radiation field.

Consider, for instance, the superposition of a left circularly polarized (LCP)
wave of frequency $\Omega$ traveling along the $z$-axis with a linearly
polarized wave of frequency $\Omega$ traveling along the $x$-axis.
Specifically, let $\chi_{ij}$ be of the form
\begin{equation}\label{chi2matrix}
{\bf \chi}(t,{\bf 0}) = \left( \begin{array}{ccc}
\cos{\Omega t} & -\sin{\Omega t} & 0 \\
-\sin{\Omega t} & 3 \cos{\Omega t} & 0 \\
0 & 0 & -4\cos{\Omega t}
\end{array} \right)
\end{equation}
up to a constant factor. It follows from (\ref{kijeq2}) that the corresponding
${\bf{\cal K}}$ is of the form
\begin{equation}
{\bf{\cal K}} = \frac{1}{2} \Omega^2 {\bf \chi}(t,{\bf 0}),
\end{equation}
so that the equation of motion (\ref{BasicEq1}) along the $z$-direction is given
by
\begin{equation} \frac{d^2z}{dt^2} + \frac{kz}{r^3} - 2\epsilon \Omega^2 z
\cos{\Omega t} = 0.
\end{equation}
This equation is satisfied by $z = 0$, which is consistent with the fact that
the orbit is always in the $(x, y)$-plane.  Therefore, to the incident radiation
field in (\ref{chimatrix}) one could add radiation fields of the type given by
(\ref{chi2matrix}).  Moreover, upon transformation to the rotating frame
\begin{equation}
{\bf{\cal K}}' = {\it R}^{-1}{\bf{\cal K}}{\it R}= \frac{1}{2}\Omega^2 \left(
\begin{array}{ccc}
\cos{\Omega t} & \sin{\Omega t} & 0 \\
\sin{\Omega t} & 3\cos{\Omega t} & 0 \\
0 & 0 & -4\cos{\Omega t}
\end{array} \right),
\end{equation}
where
\begin{equation}
{\it R} = \left(
\begin{array}{ccc}
\cos{\Omega t} & -\sin{\Omega t} & 0 \\
\sin{\Omega t} & \cos{\Omega t} & 0 \\
0 & 0 & 1
\end{array} \right)
\end{equation}
is the rotation matrix used in Section~\ref{nesec}.  
It is remarkable that the tidal
matrix ${\bf{\cal K}}'$ in the rotating frame has the same frequency $\Omega$ as
in the inertial frame; in fact,
${\bf{\cal K}}'$ can be obtained from ${\bf{\cal
K}}$ simply by letting $\Omega \rightarrow -\Omega$.

\begin{verbatim}
Original REVTEX file set up to input figures via .ps files
Fig 1:   cmrfig1a.ps     cmrfig1b.ps   cmrfig1c.ps
Fig 2:   cmrfig2a.ps     cmrfig2b.ps   cmrfig2c.ps
Fig 3:   cmrfig3a.ps     cmrfig3b.ps   cmrfig3c.ps
Fig 4:   cmrfig4a.ps     cmrfig4b.ps   cmrfig4c.ps

These lines have been set to comment statements.

If the .ps files are desired, send request  to
carmen@chicone.math.missouri.edu
and give an ftp site (the files are too big for e-mail).
Or request hard copy. 
\end{verbatim}

\begin{figure}[htb]
\caption[]{
Orbital radius versus time plots for the dynamical system given
by Hamiltonian~(\ref{Ham1}) with
initial conditions $(p_r,p_\theta,r,\theta)$
equal to $(.5, 1 , 1, 0)$ and parameter values
$\epsilon=.001$, $\delta=0$,
$\alpha=2.5$, $\beta=2$, $k=1$, and, from top to bottom,
$\Omega=2$, $\Omega=9\sqrt{3}/8$, $\Omega=1$.
\label{fig1}}
\end{figure}
\begin{figure}[htb]
\caption[]{
Orbital radius versus time plots for the dynamical system given by
the Hamiltonian~(\ref{Ham1})  
with initial conditions $(p_r,p_\theta,r,\theta)$
equal to $(.5, 1 , 1, 0)$ and parameter values
$\epsilon=.001$, $\delta=0.5$,
$\alpha=2.5$, $\beta=2$, $k=1$, and, from top to bottom,
$\Omega=2$, $\Omega=9\sqrt{3}/8$, $\Omega=1$.
\label{fig2}}
\end{figure}
\begin{figure}[htb]
\caption[]{
Orbital radius versus time plots for the dynamical system given by
the Hamiltonian~(\ref{Ham1})
with initial conditions $(p_r,p_\theta,r,\theta)$
equal to $(.5, 1 , 1, 0)$ and parameter values
$\epsilon=.001$, $\delta=1$,
$\alpha=2.5$, $\beta=2$, $k=1$, and, from top to bottom,
$\Omega=2$, $\Omega=9\sqrt{3}/8$, $\Omega=1$.
\label{fig3}}
\end{figure}
\begin{figure}[htb]
\caption[]{
Orbital radius versus time plots for the dynamical system given by
the Hamiltonian~(\ref{Ham1})
with initial conditions $(p_r,p_\theta,r,\theta)$
equal to $(.5, 1 , 1, 0)$ and parameter values
$\epsilon=.001$, $\delta=0.5$,
$\alpha=2.5$, $\beta=2$, $k=1$, $\Omega=2$.
This is an extended form of the plot presented in the top panel
of Figure~\ref{fig3}.
\label{fig4}}
\end{figure}
\end{document}